\documentclass[10pt,twocolumn]{article}
\usepackage[letterpaper,margin=0.72in,columnsep=0.24in]{geometry}
\usepackage[T1]{fontenc}
\usepackage{lmodern}
\usepackage{amsmath,amssymb,amsfonts,mathtools,amsthm,bm}
\usepackage{booktabs,array,tabularx,enumitem,float,graphicx,microtype}
\usepackage{tikz,pgfplots}
\usetikzlibrary{arrows.meta,positioning,calc}
\pgfplotsset{compat=1.18}
\usepackage{xcolor,fancyhdr,xurl,xspace,cite}
\usepackage[hidelinks]{hyperref}
\usepackage[nameinlink]{cleveref}
\setlist{nosep,leftmargin=*}
\newcolumntype{Y}{>{\centering\arraybackslash}X}
\numberwithin{equation}{section}
\newtheorem{theorem}{Theorem}[section]
\newtheorem{corollary}[theorem]{Corollary}
\newtheorem{proposition}[theorem]{Proposition}

\theoremstyle{definition}\newtheorem{definition}[theorem]{Definition}

\theoremstyle{remark}
\newcommand{\C}{\mathbb C}\newcommand{\R}{\mathbb R}
\newcommand{\N}{\mathbb N}
\newcommand{\Z}{\mathbb Z}\newcommand{\WP}{\mathbb{WP}}
\newcommand{\Tr}{\operatorname{Tr}}\newcommand{\TV}{\operatorname{TV}}
\newcommand{\ket}[1]{\left\lvert#1\right\rangle}

\title{\vspace{-1.8em}\textbf{Certified Quotient Calibration with\\
Weighted-Projective Orbits~and~Finite-Shot Guarantees}}
\author{Gunhee Cho\thanks{Corresponding author.}\\[-.3em]
\small Texas State University, San Marcos, USA
\and Jessie Wang\thanks{Jessie Wang and Angela Yue contributed equally and share second authorship.}\\[-.3em]
\small Massachusetts Institute of Technology, Cambridge, USA
\and Angela Yue\footnotemark[2]\\[-.3em]
\small Texas A\&M University, College Station, USA}
\date{August 29, 2026}

\begin{document}
\maketitle
\thispagestyle{fancy}

\begin{abstract}
Periodic controls can make a quantum-calibration scan redundant, but
compilation, noise, and measurement can invalidate the symmetry that motivates
the reduction. We introduce \emph{Certified Quotient Calibration} (CQC), which
audits whether circuit amplitudes satisfy a common positive grading and
thereby define weighted-projective-space (WPS) control orbits. From pilot
counts, it bounds the largest proposed within-orbit and smallest audited
between-orbit squared Hellinger distances by $U_{\rm in}$ and $L_{\rm out}$.
CQC accepts the quotient only when these bounds meet a prespecified task tolerance,
preserve the audited alternatives, and imply a positive net saving after
certification cost; otherwise it rejects or defers the reduction.

On the same confidence event, the certification theorem bounds every accepted
orbit substitution, preserves every prespecified competing orbit, and limits
the change of any prespecified $[0,1]$-valued calibration objective. Its cost
corollary authorizes a representative-only scan only when its
certification-inclusive cost is below the full-grid cost. Weighted-projective
descent, Hellinger testing rates, and multinomial concentration are standard;
the new result is their combination into one finite-shot approve--reject--defer
guarantee for task error and net calibration cost.

Gate-level experiments on IQM Garnet and IBM Marrakesh, Kingston, and Fez
exercise all three decisions in compiled one-, two-, and three-qubit families.
In a preregistered two-qubit Bell-phase experiment on IBM Kingston, the pilot
audit certified
$U_{\rm in}=0.01922<L_{\rm out}=0.09030$, and the quotient and full scans
selected the same correction orbit for both injected phase offsets. On fresh
held-out counts, both quotient choices satisfied the prespecified $0.03$
noninferiority margin.
After charging 18,432 pilot shots, the quotient procedure used 36,864 rather
than 55,296 total shots, a 33.3\% reduction. These results support the complete
CQC decision path for the tested families, but not a native cubic pulse law,
a device-independent WPS orbit, or a universal saving rate.
\end{abstract}


\setcounter{tocdepth}{1}
\begingroup
\small
\tableofcontents
\endgroup

\section{Introduction}\label{sec:introduction}
Calibration is a recurring cost in superconducting quantum computing.  Flux
crosstalk and uncertain coupling parameters require repeated control scans,
while periodic device responses can make part of those scans redundant
\cite{Dai2021,Bejanin2021}.  Integer phase multiplication and integer-order
responses occur in repeated-pass phase estimation and robust gate calibration
\cite{Higgins2007,Kimmel2015}, Fourier descriptions of parameterized circuits
and quantum signal processing \cite{Schuld2021,Dong2022}, third-subharmonic
transmon control \cite{Xia2026}, and circuits with rotation or Abelian symmetry
\cite{Grimsmo2020,Marvian2024}.  When the final observed response depends only
on $e^{ik\theta}$, its local phase variation can grow with $k$, whereas
$\theta$ and $\theta+2\pi j/k$ are globally indistinguishable.  Failing to
recognize a valid identification repeats physically redundant settings and
wastes shots; imposing an identification that compilation, noise, or
measurement has broken can discard distinct controls and corrupt the
calibration result.

This paper asks the following question:
\begin{quote}
Under what conditions does a winding-induced weighted-projective control orbit
remain valid after compilation, noise, and measurement, and when can orbit
representatives replace a full calibration grid while controlling task error
and yielding a positive net saving after certification cost?
\end{quote}
Answering this question requires both a structural model and a hardware-level
decision rule. The circuit model must specify which controls are predicted to
be physically equivalent, while finite-count data must determine whether the
final measurement distributions preserve those equivalences after
implementation. A Fourier peak alone does not establish the common positive
grading required for a weighted-projective quotient, and a formally valid
quotient need not survive compilation, noise, and measurement. We therefore
separate the structural proposal of candidate WPS orbits from their
simultaneous statistical certification on the implemented device.

Weighted-projective geometry supplies the structural model. Suppose that every
occupied circuit amplitude $f_x(z_0,\ldots,z_m)$ satisfies the same
positive-grading law
\begin{equation}\label{eq:positive-grading-intro}
 \begin{aligned}
 f_x(\lambda^{w_0}z_0,\ldots,\lambda^{w_m}z_m)
 &=\lambda^D f_x(z_0,\ldots,z_m),\\[-.2em]
 &\hspace{-5em}w_i,D\in\N.
 \end{aligned}
\end{equation}
The associated ray-valued state map is then constant on weighted
$\C^\times$-orbits and descends to $\WP(w_0,\ldots,w_m)$
\cite{Dolgachev1982,AbramovichHassett2011}. On a phase chart, a finite
stabilizer can induce candidate identifications of the form
$\theta\sim\theta+2\pi j/k$. Thus, WPS describes a quotient of the control
variables rather than a replacement for projective Hilbert space; when the
common grading condition fails, CQC does not assign a WPS quotient.

After compilation, noise, and measurement, the implemented circuit produces
the probability law
\begin{equation}\label{eq:final-law-intro}
 p_\theta(y)=\Tr\!\left[M_y\,\mathcal E
 \bigl(\mathcal C_\theta(\rho_0)\bigr)\right],
 \qquad y\in\mathcal Y,
\end{equation}
where $\mathcal C_\theta$ is the compiled circuit, $\mathcal E$ is the
implemented noise channel, and $\{M_y\}_{y\in\mathcal Y}$ is the fixed
measurement. We compare two such laws using the squared Hellinger distance
\begin{equation}\label{eq:h2-intro}
 H^2(P,Q)=1-\sum_{y\in\mathcal Y}\sqrt{P(y)Q(y)}
\end{equation}
Its affinity tensorizes under independent shots, so the classical
binary-testing cost has order
\begin{equation}\label{eq:testing-rate-intro}
 N=\Theta\!\left(\frac{\log(1/\delta)}{H^2(P,Q)}\right)
\end{equation}
under the stated regularity conditions
\cite{LeCamYang2000,Suresh2020,Pensia2024}. CQC does not claim this testing
rate as new: WPS specifies which orbit substitutions should be audited, while
one simultaneous confidence event and a task-aware cost rule determine
whether those substitutions may safely be used.

Weighted-projective descent, Hellinger tensorization and data processing,
multinomial concentration, and total-variation bounds for bounded observables
are standard ingredients. The new result is their joint finite-sample
conclusion. Under an audited positive grading and one simultaneous confidence
event, \Cref{thm:certified-quotient} certifies every proposed within-orbit
substitution, preserves every prespecified between-orbit distinction, and
bounds the change of any prespecified $[0,1]$-valued calibration objective. Its
cost consequence, \Cref{cor:net-saving}, authorizes the reduced scan
only when
\begin{equation}\label{eq:net-saving-intro}
 N_{\rm cert}+\lvert\Theta_M/G\rvert n_{\rm cal}<M n_{\rm cal}.
\end{equation}
In the resulting procedure, the verified grading first determines candidate
WPS control orbits.
Simultaneous confidence bounds $U_{\rm in}$ and $L_{\rm out}$ then determine
whether the proposed identifications are accepted, rejected, or deferred. An
accepted quotient is used for calibration only if its total cost, including
certification, is smaller than that of the full scan.

Certification is not free. On the prespecified Bernoulli-parity submodel used
in the IBM experiments, \Cref{thm:audit-cost} shows that resolving a fixed
pair separated by squared Hellinger gap $\Delta$ with error probability at
most $\delta$ requires order $\Delta^{-1}\log(1/\delta)$ samples in the worst
case, and it provides a simultaneous upper bound with only a logarithmic
multiplicity penalty for the audited family. Consequently, a nonreusable
certificate must be charged against the settings removed by the quotient. If
the same certificate remains valid for $T$ calibration scans within one
monitored epoch, the quotient becomes amortized shot-saving precisely when
\begin{equation}\label{eq:amortized-intro}
 T>\frac{N_{\rm cert}}
 {(M-\lvert\Theta_M/G\rvert)n_{\rm cal}}.
\end{equation}
This result prevents a reduction in the number of calibration settings from
being reported as a net saving when the audit required to justify that
reduction costs more than it saves.

The hardware study tests the decisions produced by this framework rather than
claiming to validate the full theory. IQM Garnet provides the initial
gate-level realization of programmed $k=1$ and $k=3$ families. IBM Marrakesh,
Kingston, and Fez extend the audit to entangled two- and three-qubit circuits
and independent shot budgets: strong programmed $k=3$ families test
acceptance, $k=1$ controls test rejection of a false threefold quotient, and
weak $k=3$ families test deferral. A preregistered Kingston experiment then
executes the complete CQC procedure using pilot certification, independent
exhaustive and quotient scans, two prespecified phase offsets, and fresh
held-out Bell-loss evaluation. Circuit order and analysis rules were fixed
before the outcomes were examined, and no recorded stage was selected or
resubmitted in response to its result.

The cross-platform claim is procedural rather than numerical. The same grading
audit, simultaneous confidence rule, and net-cost test can be applied when the
platform, circuit size, signal strength, or shot budget changes, but
device-specific Hellinger gaps are analyzed separately and the observed saving
rate is not treated as universal. The Kingston experiment tests preservation
of the prespecified calibration objective within one specified backend, physical
layout, and calibration epoch. The remaining experiments test whether the
accept, reject, and defer decisions can be reproduced under other compiled
measurement stacks. Repeating the same audit in these settings shows that the
procedure can be executed on the tested platforms. It does not establish
equality of their noise channels or a device-independent performance guarantee.

Existing methods address the local sensitivity, spectral structure, and
repetition schedules of parameterized quantum circuits separately. Quantum
Fisher information gives the largest local sensitivity available before
fixing a measurement \cite{BraunsteinCaves1994}, but it does not determine
whether distant controls produce the same final observed law. Fourier analyses
identify frequencies accessible to parameterized circuits
\cite{Schuld2021,Casas2023}, yet the presence of an integer harmonic alone does
not imply the common positive grading required by
\eqref{eq:positive-grading-intro}. Robust phase-estimation and calibration
methods design repetition schedules or infer phase parameters efficiently
\cite{Higgins2007,Kimmel2015,Dong2025}. CQC instead begins with a proposed
control quotient and tests whether it remains valid after a specified
compilation, noise channel, and measurement.
Hellinger distance has also been used to assess the reproducibility of noisy
quantum circuits \cite{DasguptaHumble2022}, and classical statistics supplies
the testing and concentration results used in the audit
\cite{LeCamYang2000,Suresh2020,Pensia2024,Weissman2003}. These results do not
specify which control settings should be identified by a weighted-projective
orbit, nor do they determine whether replacing those settings by orbit
representatives is safe and cost-effective. The contribution of CQC is the
following combined guarantee: a structurally justified candidate control orbit is tested by one
simultaneous finite-sample certificate that covers both proposed substitutions
and prespecified competitors, bounds the prespecified task error, and charges the
audit before authorizing a reduced calibration scan. No novelty is claimed for
weighted-projective descent, the Hellinger testing rate, or multinomial
concentration in isolation.

The remainder of the paper is organized as follows. \Cref{sec:model} defines
the compiled finite-count model, the weighted-projective control quotient, and
a running calibration example. \Cref{sec:theory} proves the simultaneous
quotient certificate, the net-saving criterion, and the audit-cost bound.
\Cref{sec:algorithm-experiment} specifies the CQC algorithm and reports its
mechanism, safety, and end-to-end hardware tests. \Cref{sec:discussion}
discusses applications, limitations, and reproducibility.

\section{Model and running calibration example}\label{sec:model}

\subsection{A reader's map of the three layers}
Section~\ref{sec:model} separates three questions that are easy to conflate.
The \emph{physical layer} asks which probability law is produced by a
compiled circuit, a fixed noise channel, and a chosen measurement. The
\emph{structural layer} asks whether the circuit amplitudes possess a positive
grading and, if so, which control settings the resulting weighted-projective
action proposes to identify. The \emph{statistical layer} asks whether finite
hardware counts support those proposed identifications while still separating
the competing orbits relevant to calibration. These layers have different
roles: the first defines the observed experiment, the second proposes
a quotient, and only the third can authorize its use on the audited hardware.

The notation follows this division. The family $p_\theta$ denotes the unknown
distribution of one recorded shot at control setting $\theta$;
$C_\theta$ and $\widehat p_\theta$ are its observed count vector and empirical
distribution. A finite group $G$ acts on the requested grid $\Theta_M$ and
defines candidate control orbits $[\theta]_G$. Their statistical quality is summarized
by the largest within-orbit distance $h_{\rm in}^2$ and the smallest
prespecified between-orbit distance $h_{\rm out}^2$. The computable bounds
$U_{\rm in}$ and $L_{\rm out}$ are, respectively, an upper confidence bound
for the former and a lower confidence bound for the latter. Thus the reader
may interpret the symbols in the order \emph{observed counts, candidate
control orbit, simultaneous confidence bounds, decision}, without assuming in advance that the
structural quotient survives compilation and noise.

The ingredients themselves are standard but serve distinct purposes here.
Weighted-projective descent and its stabilizers belong to weighted-projective
geometry \cite{Dolgachev1982,AbramovichHassett2011}; statistical experiments,
data processing, and finite-sample discrimination belong to classical
decision theory and information theory \cite{LeCamYang2000,CoverThomas2006,
Suresh2020,Pensia2024}; and quantum Fisher information bounds the local
information available before a particular measurement is fixed
\cite{BraunsteinCaves1994}. The contribution developed in
Sections~\ref{sec:theory}--\ref{sec:algorithm-experiment} is the decision rule
that connects these pieces: a proposed WPS control orbit is accepted only through a
simultaneous finite-count certificate and is used only when the
certification-inclusive calibration cost decreases.

\subsection{From a compiled circuit to finite counts}
Let $\mathcal H$ be a finite-dimensional complex Hilbert space, let $\Theta$
be a compact control space, and let $\Theta_M\subset\Theta$ be the finite grid
requested for calibration. For each $\theta\in\Theta_M$, the circuit map
$\mathcal C_\theta$ acts on an initial density operator $\rho_0$ on
$\mathcal H$. During one calibration epoch---an interval over which the
residual channel and measurement are treated as fixed---a parameter-independent
completely positive trace-preserving (CPTP) map $\mathcal E$ represents the
compiled residual hardware channel, and $\{M_y:y\in\mathcal Y\}$ is a fixed
positive-operator-valued measure (POVM). The observed law is
\begin{equation}\label{eq:observed-law-model}
 p_\theta(y)=\Tr\!\left[M_y\,\mathcal E
 \bigl(\mathcal C_\theta(\rho_0)\bigr)\right].
\end{equation}
Here parameter independence means that the same $\mathcal E$ and POVM are used
for every audited $\theta$ within the epoch; detected control dependence or
drift requires recertification.
When several measurement bases are used, let $s\in\mathcal S$ denote the
recorded basis label, let $\pi_s>0$ with
$\sum_{s\in\mathcal S}\pi_s=1$ be its prespecified design weight, and let
$p_{s,\theta}$ be the outcome law conditional on that basis. The corresponding
joint law satisfies
\begin{equation}\label{eq:joint-design-law}
 \begin{aligned}
 p_\theta(s,y)&=\pi_s p_{s,\theta}(y),\\
 H^2(p_\theta,p_{\theta'})
 &=\sum_{s\in\mathcal S}\pi_s
 H^2(p_{s,\theta},p_{s,\theta'}).
 \end{aligned}
\end{equation}
The theorems may therefore be applied to this single joint distribution. For
$S$ equally weighted bases, the unweighted sum of the basiswise squared
Hellinger distances is $S$ times the distance of the joint law and must be
divided by $S$ before comparison with a probability-distance threshold. Actual
randomization of the basis label is unnecessary: in a fixed stratified design,
one constructs a confidence region for each $(\theta,s)$ law, allocates the
total error probability across those regions, and combines their bounds with
the weights $\pi_s$.

At each requested setting $\theta$, suppose that $n_\theta$ shots are
conditionally independent within the calibration epoch and satisfy
\begin{equation}\label{eq:categorical}
 Y_{\theta,i}\overset{\mathrm{iid}}{\sim}
 \operatorname{Categorical}(p_\theta),
 \qquad i=1,\ldots,n_\theta.
\end{equation}
Here $Y_{\theta,i}$ is the outcome of one shot. The count vector and empirical
distribution are
\begin{equation}\label{eq:counts-from-categorical}
 \begin{gathered}
 C_\theta(y)=\sum_{i=1}^{n_\theta}{\bf1}\{Y_{\theta,i}=y\},
 \qquad C_\theta\sim\operatorname{Multinomial}(n_\theta,p_\theta),\\
 \widehat p_\theta(y)=\frac{C_\theta(y)}{n_\theta}.
 \end{gathered}
\end{equation}
Thus $p_\theta$ is the probability vector fixed by the calibration epoch,
whereas $Y_{\theta,i}$, $C_\theta$, and $\widehat p_\theta$ are random. Drift
or overdispersion can violate the conditional-independence model and therefore
requires a separate diagnostic or recertification.

The fixed CPTP map $\mathcal E$ may include coherent residual error,
dephasing, amplitude damping, coupling to an unobserved environment, and
compilation effects that remain stable during the audit. Classical readout
noise can be appended as a parameter-independent Markov kernel $T(z\mid y)$,
where $T(z\mid y)$ is the conditional probability of recording $z$ when the
pre-readout outcome is $y$. The recorded law is then
\begin{equation}\label{eq:readout-kernel}
 q_\theta(z)=\sum_{y\in\mathcal Y}T(z\mid y)p_\theta(y).
\end{equation}
Parameter independence means that $T$ does not vary with $\theta$ during the
epoch. Such postprocessing cannot increase Hellinger separation, although it
may weaken or erase an orbit signature. This is an epoch-level null model, not
a claim that hardware noise is universally control independent; detected
control dependence or drift requires a revised model and recertification.

Let a finite group $G$, obtained from the structural grading audit, act on the
calibration grid $\Theta_M$. The orbit of $\theta$ is
\begin{equation}\label{eq:orbit-definition}
 [\theta]_G=\{g\theta:g\in G\},
\end{equation}
and the quotient set $\Theta_M/G$ consists of all such orbits. Write
$M=\lvert\Theta_M\rvert$ for the number of full-grid settings and
$R=\lvert\Theta_M/G\rvert$ for the number of orbit representatives. The
quotient is exact at the level of the observed experiment when
\begin{equation}\label{eq:exact-invariance}
 p_{g\theta}=p_\theta
 \qquad(\theta\in\Theta_M,\ g\in G).
\end{equation}
Under exact invariance, any member of an orbit may be replaced by its
representative without changing the observed law. CQC does not assume this
equality on hardware: it treats $G$ as a structurally proposed action and uses
finite-count data to decide whether an approximate quotient is accurate enough
for the prespecified calibration task.

\begin{definition}[Within- and between-orbit gaps]\label{def:gaps}
Fix, before examining confirmatory outcomes, a set $\mathcal B$ of audited
pairs whose members belong to different candidate orbits. Define
\begin{align}
 h_{\rm in}^2
 &=\max_{\theta\in\Theta_M,\,g\in G}
 H^2(p_\theta,p_{g\theta}),\label{eq:hin}\\
 h_{\rm out}^2
 &=\min_{(\theta,\theta')\in\mathcal B}
 H^2(p_\theta,p_{\theta'}).\label{eq:hout}
\end{align}
The within-orbit quantity $h_{\rm in}^2$ is the worst distributional error
incurred by a proposed orbit substitution, whereas $h_{\rm out}^2$ is the
smallest separation among the prespecified competing orbits. Exact invariance
gives $h_{\rm in}^2=0$. In finite data, CQC instead constructs an upper
confidence bound $U_{\rm in}$ for $h_{\rm in}^2$ and a lower confidence bound
$L_{\rm out}$ for $h_{\rm out}^2$. The set $\mathcal B$ may contain every
different-orbit pair or only the competitors relevant to the declared
calibration task, but it must be fixed before the confirmatory counts are
inspected.
\end{definition}

The square-root map embeds a probability vector $p$ into the positive orthant
of the unit sphere as $\sqrt p=(\sqrt{p(y)})_{y\in\mathcal Y}$, and
\begin{equation}\label{eq:hellinger-embedding}
 H^2(P,Q)=\frac12\|\sqrt P-\sqrt Q\|_2^2
 =1-\sum_{y\in\mathcal Y}\sqrt{P(y)Q(y)}.
\end{equation}
For a differentiable full-support model, define the classical Fisher
information matrix by
\begin{equation}\label{eq:classical-fisher-definition}
 [F_C(\theta)]_{ab}
 =\sum_{y\in\mathcal Y}
 \frac{\partial_a p_\theta(y)\,\partial_b p_\theta(y)}{p_\theta(y)}.
\end{equation}
Then
\begin{equation}\label{eq:hellinger-fisher-local}
 H^2(p_\theta,p_{\theta+d\theta})
 =\frac18\,d\theta^{\mathsf T}F_C(\theta)d\theta
 +o(\|d\theta\|^2).
\end{equation}
Thus the same distance gives the local Fisher geometry, the global orbit gaps,
and the finite-shot testing scale. It remains finite when some probabilities
vanish, and a parameter-independent Markov kernel can only contract it. Such
contraction may reduce both $h_{\rm in}$ and $h_{\rm out}$; noise therefore
does not make a proposed quotient safe merely by making within-orbit
distributions look closer. The square-root embedding and its local Fisher
expansion are standard information-geometric facts; Hellinger testing theory
supplies the corresponding finite-sample discrimination scale
\cite{LeCamYang2000,Suresh2020,Pensia2024}. What must be proved later is not
this metric identity, but that simultaneous bounds on the two orbit gaps yield
a valid quotient-calibration decision and a net saving after certification.

\subsection{When a WPS orbit is available}
Let $\mathcal X$ be the finite set of occupied computational-basis labels, and
let
\begin{equation}\label{eq:polynomial-amplitude-map}
 F=(f_x)_{x\in\mathcal X}:
 \C^{m+1}\setminus\{0\}\longrightarrow\C^{\mathcal X}
\end{equation}
be a polynomial amplitude map that is not identically zero. Wherever
$F(z)\neq0$, the physical pure state is represented by the projective ray
$[F(z)]$. Divide all components by their greatest common polynomial factor;
this leaves $[F(z)]$ unchanged wherever the original ray is defined and removes
an extraneous scalar contribution to the grading. Writing
\begin{equation}\label{eq:exponent-support}
 \begin{aligned}
 f_x(z)&=\sum_{a\in\N^{m+1}}c_{x,a}z^a,\\
 \operatorname{supp}(f_x)
 &=\{a\in\N^{m+1}:c_{x,a}\neq0\}.
 \end{aligned}
\end{equation}
the existence of a positive weighted-projective quotient can be tested by a
finite system of linear equations on the occupied exponent vectors.

\begin{proposition}[Positive-grading audit]\label{prop:grading}
For the reduced polynomial amplitude map $F$, the following conditions are
equivalent:
\begin{enumerate}[label=(\roman*),leftmargin=2.2em]
\item There exist weights $w=(w_0,\ldots,w_m)\in\Z_{>0}^{m+1}$ and a degree
$D\in\Z_{\ge0}$ such that
\begin{equation}\label{eq:grading-equivariance}
 \begin{aligned}
 F(\lambda^w z)&=\lambda^D F(z),\\
 \lambda^w z&:=(\lambda^{w_0}z_0,\ldots,\lambda^{w_m}z_m),
 \qquad \lambda\in\C^\times.
 \end{aligned}
\end{equation}
\item Every occupied monomial exponent satisfies
\begin{equation}\label{eq:grading-linear}
 w\cdot a=D
 \qquad
 \bigl(a\in\operatorname{supp}(f_x),\ x\in\mathcal X\bigr).
\end{equation}
\end{enumerate}
When these conditions hold, the ray map $[F]$ is constant on weighted
$\C^\times$-orbits and descends, on its base-point-free domain, to the
weighted-projective quotient $\WP(w_0,\ldots,w_m)$.
\end{proposition}
\begin{proof}
If condition (ii) holds, then every component satisfies
\[
 f_x(\lambda^{w_0}z_0,\ldots,\lambda^{w_m}z_m)
 =\sum_a c_{x,a}\lambda^{w\cdot a}z^a
 =\lambda^D f_x(z).
\]
Hence $F(\lambda^wz)=\lambda^D F(z)$, and multiplication of the entire
amplitude vector by the nonzero scalar $\lambda^D$ leaves its projective ray
unchanged. This proves condition (i) and the descent of $[F]$.

Conversely, every algebraic character of $\C^\times$ has the form
$\lambda\mapsto\lambda^D$ for some $D\in\Z$. If
$F(\lambda^wz)=\lambda^D F(z)$, comparison of coefficients of the distinct
monomials $z^a$ gives $\lambda^{w\cdot a}=\lambda^D$ for every occupied
exponent and every $\lambda\in\C^\times$. Therefore $w\cdot a=D$. Positivity
of $w$ is a separate feasibility condition and does not follow from
periodicity alone. This is the standard weighted-projective descent criterion
\cite{Dolgachev1982,AbramovichHassett2011}.
\end{proof}

\subsection{A complete one-qubit calculation}
The calculation below follows one example from end to end: it asks whether the
$k$-fold redundancy proposed by WPS remains in the measured distribution after
noise, how many shots are sufficient to resolve it, and whether the calibration
grid can be reduced after charging the certification cost.

Consider the one-qubit family
\begin{equation}\label{eq:state-family}
 \begin{aligned}
 \ket{\psi_{k,r}(\theta)}
 &=\frac{\ket0+r^k e^{ik\theta}\ket1}{\sqrt{1+r^{2k}}},\\
 &\theta\in\mathbb T,\quad k\in\Z_{>0},\quad r>0.
 \end{aligned}
\end{equation}
where $\mathbb T=\R/(2\pi\Z)$.  Define
\begin{equation}\label{eq:contrast}
 a_k(r)=\frac{2r^k}{1+r^{2k}},\qquad
 z_k(r)=\frac{1-r^{2k}}{1+r^{2k}}.
\end{equation}
The Bloch vector of \eqref{eq:state-family} is
\begin{equation}\label{eq:bloch-vector}
 \begin{gathered}
 \bigl(a_k(r)\cos(k\theta),\,
       a_k(r)\sin(k\theta),\,z_k(r)\bigr),\\
 a_k(r)^2+z_k(r)^2=1.
 \end{gathered}
\end{equation}
Thus $k$ is the winding order: as $\theta$ traverses one period, the Bloch
vector makes $k$ azimuthal revolutions.  The scale $r$ determines both the
observable transverse contrast $a_k(r)$ and the latitude $z_k(r)$.

The homogeneous lift $F(\alpha,\beta)=(\alpha,\beta^k)$ has common degree
$k$ for weights $(k,1)$ and therefore descends to $\WP(k,1)$.  On the chart
$\alpha\ne0$, normalization to $\alpha=1$ leaves the residual action
$\beta\mapsto\zeta\beta$, $\zeta^k=1$, on the chart cover.  The candidate
identification proposed by this lift is therefore
\begin{equation}\label{eq:k-orbit}
 \theta\sim\theta+\frac{2\pi j}{k},\qquad j=0,\ldots,k-1.
\end{equation}
This is not a direct measurement of a generic stack stabilizer.  It records
the $k$ control representations on the chosen weighted chart cover that map
to one quotient coordinate.

The same normalized state determines a point of $\mathbb{CP}^1$, equivalently
a point on the Bloch sphere.  The Bloch point completely specifies the prepared
one-qubit pure state at a fixed control value, but it does not retain the
provenance of its parametrization.  It therefore cannot by itself distinguish
lifts with different $k$ or recover the $k$ control representations in
\eqref{eq:k-orbit}.  The WPS lift instead retains the orbit identification on
the chosen chart cover and its ramification at $\beta=0$.  The residual group
fixes that collapsed point, whereas the stack stabilizer is trivial at a
generic point of $\WP(k,1)$ with both coordinates nonzero.  Thus WPS does not
replace the Bloch state space; it adds the control-space quotient structure
that the Bloch representation forgets.

For a differentiable pure-state family, the quantum Fisher information is
\[
 F_Q(\theta)=4\left(
 \langle\partial_\theta\psi\mid\partial_\theta\psi\rangle
 -\left|\langle\psi\mid\partial_\theta\psi\rangle\right|^2
 \right).
\]
Direct substitution of \eqref{eq:state-family} gives
\begin{equation}\label{eq:qfi-running}
 F_Q(k,r)=4k^2\frac{r^{2k}}{(1+r^{2k})^2}=k^2a_k(r)^2.
\end{equation}
In this one-parameter pure-state model, this is the measurement-independent
upper bound on local information and is locally attainable by a suitable
measurement.  The factor $k^2$ expresses phase amplification, whereas
$a_k(r)^2$ limits the available contrast; a large $k$ alone therefore does not
guarantee high sensitivity.  Moreover, QFI is a local differential quantity
and does not record the global identification \eqref{eq:k-orbit}.  The WPS lift
proposes that identification, while the implemented classical Fisher
information and Hellinger gaps determine whether it remains observable after
the selected channel and measurement.

Let the phase-covariant coherence-contraction channel be
\begin{equation}\label{eq:dephasing-channel}
 \begin{aligned}
 \rho&=\begin{pmatrix}\rho_{00}&\rho_{01}\\ \rho_{10}&\rho_{11}\end{pmatrix},\\
 \mathcal D_\eta(\rho)
 &=\begin{pmatrix}\rho_{00}&\eta\rho_{01}\\
                   \eta\rho_{10}&\rho_{11}\end{pmatrix},
 \qquad 0\le\eta\le1.
 \end{aligned}
\end{equation}
Here $\eta$ is the fraction of coherence retained after noise, rather than the
noise strength itself.  Fixed $X$, $Y$, and $Z$ measurements after this channel
give
\begin{align}
 p_{X,\theta}(\pm)&=\frac{1\pm\eta a_k(r)\cos(k\theta)}2,\label{eq:px}\\
 p_{Y,\theta}(\pm)&=\frac{1\pm\eta a_k(r)\sin(k\theta)}2,\label{eq:py}\\
 p_{Z,\theta}(\pm)&=\frac{1\pm z_k(r)}2.\label{eq:pz}
\end{align}
Thus this illustrative channel preserves the orbit identification
\eqref{eq:k-orbit} exactly while reducing the observable transverse contrast
from $a_k(r)$ to $\eta a_k(r)$.  A general compilation, noise, and measurement
pipeline need not preserve that identification.  CQC therefore tests the
equality of the observed laws rather than inferring it from the structural
candidate relation alone.

For the binary law $b(u)=((1+u)/2,(1-u)/2)$, define
\begin{equation}\label{eq:binary-h2}
 \begin{aligned}
 h(u,v)&:=H^2(b(u),b(v))\\
 &=1-\frac12\bigl(\sqrt{(1+u)(1+v)}
 +\sqrt{(1-u)(1-v)}\bigr).
 \end{aligned}
\end{equation}
Set $c_\theta=\eta a_k(r)\cos(k\theta)$ and
$s_\theta=\eta a_k(r)\sin(k\theta)$.  The exact setting-summed Hellinger
separation is
\begin{equation}\label{eq:exact-xy-h2}
 H^2_{XY}(\theta,\theta+\Delta)
 =h(c_\theta,c_{\theta+\Delta})
 +h(s_\theta,s_{\theta+\Delta}).
\end{equation}
In the weak-contrast regime $\eta a_k(r)\ll1$, this becomes
\begin{equation}\label{eq:weak-h2}
 H^2_{XY}(\theta,\theta+\Delta)
 =\frac{\eta^2a_k(r)^2}{4}
 \bigl[1-\cos(k\Delta)\bigr]
 +O\!\left(\eta^4a_k(r)^4\right),
\end{equation}
For $\eta a_k(r)>0$, the exact expression \eqref{eq:exact-xy-h2} vanishes if
and only if $k\Delta\in2\pi\Z$, precisely the candidate identification in
\eqref{eq:k-orbit}.  If $\eta a_k(r)=0$, every phase candidate has the same
observed law and no number of shots can certify the winding.  The approximation
also shows why a large nominal $k$ does not guarantee a cheap certificate: at
fixed $r<1$, the contrast $a_k(r)$ may collapse faster than the phase factor
grows.

The implemented $X$- and $Y$-measurement Fisher informations are
\begin{align}
 F_X(\theta)&=\frac{\eta^2a_k(r)^2k^2\sin^2(k\theta)}
 {1-\eta^2a_k(r)^2\cos^2(k\theta)},\label{eq:fx-running}\\
 F_Y(\theta)&=\frac{\eta^2a_k(r)^2k^2\cos^2(k\theta)}
 {1-\eta^2a_k(r)^2\sin^2(k\theta)}.\label{eq:fy-running}
\end{align}
They quantify local sensitivity after the chosen channel and measurement.
The local expansion of the exact Hellinger separation is
\begin{equation}\label{eq:local-fisher-h2}
 H^2_{XY}(\theta,\theta+\Delta)
 =\frac{\Delta^2}{8}\bigl(F_X(\theta)+F_Y(\theta)\bigr)
 +o(\Delta^2).
\end{equation}
If shots are divided equally between $X$ and $Y$, the implemented Fisher
information per shot is $(F_X+F_Y)/2$.  Thus $F_Q$ is the premeasurement local
upper bound, $F_X$ and $F_Y$ are the local information retained by the selected
channel and measurements, and Hellinger separation extends the comparison to
finite control displacements.  The factor $k$ amplifies phase sensitivity,
while the surviving contrast $\eta a_k(r)$ and the global period $2\pi/k$
determine whether that amplification is useful for calibration.

For two simple hypotheses $P$ and $Q$ with equal prior probabilities, the
Hellinger affinity tensorizes as
\begin{equation}\label{eq:affinity-tensorization}
 \operatorname{Aff}(P^{\otimes N},Q^{\otimes N})
 =\bigl(1-H^2(P,Q)\bigr)^N.
\end{equation}
The Bhattacharyya error bound therefore gives the affinity-based sufficient
budget
\begin{equation}\label{eq:known-shot-order}
 N_{\rm suff}(P,Q,\delta)=
 \left\lceil
 \frac{\log(1/(2\delta))}
 {-\log\!\left(1-H^2(P,Q)\right)}
 \right\rceil .
\end{equation}
When $H^2(P,Q)\ll1$, this is asymptotic to
$\log(1/(2\delta))/H^2(P,Q)$.  This is a sufficient budget for one
prespecified pair, not the full CQC sample size.  CQC must estimate many
within- and between-orbit distances simultaneously, so $N_{\rm cert}$ is set
by the confidence construction that produces $U_{\rm in}$ and $L_{\rm out}$.

Let $\Theta_M$ be a $G$-invariant calibration grid with
$M=|\Theta_M|$ settings, and let
\begin{equation}\label{eq:representative-count-example}
 R=|\Theta_M/G|
\end{equation}
be the number of orbit representatives.  If $G\simeq\Z_k$ acts freely on the
grid and every orbit is complete, then $R=M/k$.  With $n_{\rm cal}$ shots per
setting, the full scan costs $Mn_{\rm cal}$ shots, whereas the certified
quotient scan costs $N_{\rm cert}+Rn_{\rm cal}$.  Conditional on approval of
the quotient, the necessary and sufficient condition for a net shot saving is
\begin{equation}\label{eq:net-saving-example}
 N_{\rm cert}<(M-R)n_{\rm cal}.
\end{equation}
The cost test is reached only when $U_{\rm in}$ is below the prespecified
within-orbit tolerance, $L_{\rm out}$ exceeds the prespecified separation
requirement, and the confidence bounds resolve an accept, reject, or defer
decision. This example shows the role of each component: WPS proposes
which settings form an orbit, Hellinger certification determines whether that
grouping is admissible, and the final cost test determines whether the reduced
scan should be used in calibration.

\begin{figure*}[t]
\centering
\begin{tikzpicture}[>=Latex,node distance=8mm and 10mm,
box/.style={draw,rounded corners,align=center,minimum height=10mm,
text width=28mm,font=\small},
lab/.style={font=\scriptsize,fill=white,inner sep=1pt}]
\node[box] (g) {positive grading\\candidate WPS orbit};
\node[box,right=of g] (p) {pilot counts\\on audited pairs};
\node[box,right=of p] (h) {$U_{\rm in}$ and\\$L_{\rm out}$};
\node[box,right=of h] (d) {accept, reject,\\or defer};
\node[box,below=of d] (c) {certification-inclusive\\net-cost test};
\node[box,below=of c] (q) {quotient\\calibration scan};
\node[box,below=of h] (f) {retain full grid\\or collect more data};
\draw[->] (g)--(p);\draw[->] (p)--(h);\draw[->] (h)--(d);
\draw[->] (d)--node[lab,right]{accept}(c);
\draw[->] (d.south west)--node[lab,pos=.34,left]{reject/defer}(f.north east);
\draw[->] (c)--node[lab,right]{net saving}(q);
\draw[->] (c.west)--node[lab,below]{no net saving}(f.east);
\end{tikzpicture}
\caption{CQC separates a structural candidate orbit from equality of the
observed laws.  An accepted candidate proceeds to quotient calibration only
after passing the certification-inclusive net-cost test; rejection, deferral,
or failure to save retains the full grid or requests more data.}
\label{fig:pipeline}
\end{figure*}

\section{Certified quotient calibration}\label{sec:theory}

\subsection{Simultaneous certificate}
Let $\widehat p_\theta$ denote the empirical distribution from the
$n_\theta$ shots collected at setting $\theta$, and let $p_\theta$ denote
the corresponding unknown true observed law.  A confidence procedure returns
data-dependent bounds $U_{\rm in}$ and $L_{\rm out}$ such that, for every
data-generating family in its stated coverage class,
\begin{equation}\label{eq:coverage}
 \Pr_{\{p_\theta\}}\left\{h_{\rm in}^2\le U_{\rm in}
 \ \text{and}\ h_{\rm out}^2\ge L_{\rm out}\right\}\ge1-\delta.
\end{equation}
The probability is over repeated sampling of all audited count records,
whereas $h_{\rm in}$ and $h_{\rm out}$ are fixed functionals of the true
observed laws.  The construction may use exact multinomial regions, a
prespecified parametric likelihood with valid coverage, or a bootstrap whose
coverage has been established for the declared model; the theorem requires
only the joint event \eqref{eq:coverage}.

The simultaneous event can be constructed directly from the observed counts.
Let $M=|\Theta_M|$, let $d=|\mathcal Y|$, and assign failure probability
$\delta/M$ to each audited setting. Define
\begin{equation}\label{eq:multinomial-radius}
 \epsilon_\theta=\min\!\left\{2,
 \sqrt{\frac{2}{n_\theta}
 \log\frac{(2^d-2)M}{\delta}}\right\},
 \qquad r_\theta=\sqrt{\epsilon_\theta/2}.
\end{equation}
For each audited pair $(\theta,\theta')$, define
\begin{equation}\label{eq:pair-bounds}
 \begin{aligned}
 U_{\theta,\theta'}&=\min\{1,[H(\widehat p_\theta,
 \widehat p_{\theta'})+r_\theta+r_{\theta'}]^2\},\\
 L_{\theta,\theta'}&=[H(\widehat p_\theta,
 \widehat p_{\theta'})-r_\theta-r_{\theta'}]_+^2.
 \end{aligned}
\end{equation}
where $[x]_+=\max\{x,0\}$. Set
\[
 U_{\rm in}=\max_{\theta\in\Theta_M,\,g\in G}U_{\theta,g\theta},
 \qquad
 L_{\rm out}=\min_{(\theta,\theta')\in\mathcal B}L_{\theta,\theta'}.
\]

\begin{theorem}[Finite-sample simultaneous multinomial certificate]
\label{thm:explicit-multinomial}
Under the categorical sampling model \eqref{eq:counts-from-categorical},
suppose that $\Theta_M$, $G$, $\mathcal B$, and the shot allocations
$\{n_\theta\}_{\theta\in\Theta_M}$ are fixed before the audited counts are
inspected. The aggregate bounds constructed from
\eqref{eq:multinomial-radius}--\eqref{eq:pair-bounds} satisfy
\eqref{eq:coverage}. The guarantee is nonasymptotic, permits unequal
$n_\theta$, and simultaneously covers every proposed within-orbit
substitution and every pair in $\mathcal B$. Independence among the
settingwise confidence events is not required.
\end{theorem}
\begin{proof}
For each $\theta$, the multinomial deviation inequality of Weissman
et al.~\cite{Weissman2003} gives, for $0<\epsilon<2$,
\[
 \Pr_{\{p_\theta\}}\!\left\{
 \|\widehat p_\theta-p_\theta\|_1>\epsilon\right\}
 \le (2^d-2)e^{-n_\theta\epsilon^2/2}.
\]
If the square-root term in \eqref{eq:multinomial-radius} is smaller than
$2$, substituting $\epsilon=\epsilon_\theta$ makes the right-hand side
$\delta/M$. If the truncation is active, then $\epsilon_\theta=2$, and the
event has probability zero because the $L^1$ distance between probability
distributions cannot exceed $2$. Hence, in either case,
\[
 \Pr_{\{p_\theta\}}\!\left\{
 \|\widehat p_\theta-p_\theta\|_1>\epsilon_\theta\right\}
 \le\delta/M.
\]
A union bound therefore gives an event $\mathcal E$ of probability at least
$1-\delta$ on which all $M$ settingwise deviation bounds hold; no
independence among these events is used. Since
\[
 H^2(P,Q)\le\TV(P,Q)=\frac12\|P-Q\|_1,
\]
the event $\mathcal E$ implies
$H(\widehat p_\theta,p_\theta)\le r_\theta$ for every setting. Hence, for
every audited pair,
\[
 \bigl[H(\widehat p_\theta,\widehat p_{\theta'})
 -r_\theta-r_{\theta'}\bigr]_+
 \le H(p_\theta,p_{\theta'})
 \le H(\widehat p_\theta,\widehat p_{\theta'})
 +r_\theta+r_{\theta'}.
\]
Squaring these inequalities and using $H^2\le1$ gives
$L_{\theta,\theta'}\le H^2(p_\theta,p_{\theta'})\le
U_{\theta,\theta'}$. Maximizing over candidate-orbit substitutions and
minimizing over $\mathcal B$ proves \eqref{eq:coverage}.
\end{proof}

The radius in \eqref{eq:multinomial-radius} is a conservative closed-form
choice that can be recomputed from the frozen counts, $M$, $d$, and $\delta$.
It may be replaced by any procedure satisfying the joint coverage requirement
\eqref{eq:coverage}, including exact multinomial regions or valid confidence
sequences. If sampling continues after a defer decision, fixed-$n$ bounds
cannot be reused at successive looks without adjusting their coverage. The
protocol must instead prespecify an error-spending rule or use a time-uniform
confidence sequence.

Fix a declared within-orbit tolerance $\varepsilon_{\rm orb}$ and a
separation margin $\gamma>0$. CQC accepts when
\begin{equation}\label{eq:accept-rule}
 U_{\rm in}\le\varepsilon_{\rm orb},
 \qquad L_{\rm out}-U_{\rm in}\ge\gamma.
\end{equation}
The first inequality bounds the worst proposed substitution error, while the
second ensures that every competitor in $\mathcal B$ remains separated from
the candidate within-orbit discrepancies by at least $\gamma$. CQC rejects
when
\[
 \max_{\theta\in\Theta_M,\,g\in G}L_{\theta,g\theta}
 >\varepsilon_{\rm orb},
\]
because at least one proposed substitution is then certified to exceed the
tolerance. In all remaining cases, CQC returns defer.

\begin{theorem}[Certified weighted-projective quotient]\label{thm:certified-quotient}
Assume that (i) the amplitude model satisfies \cref{prop:grading} and induces
a finite candidate action $G$ on $\Theta_M$; (ii) the calibration epoch is
governed by the fixed observed laws \eqref{eq:final-law-intro};
(iii) $\mathcal B$ and the confidence procedure are fixed before confirmatory
outcomes are inspected; and (iv)
\eqref{eq:coverage} holds, for example by
\cref{thm:explicit-multinomial}. If CQC accepts, then with probability at least
$1-\delta$:
\begin{enumerate}[label=(\alph*),leftmargin=2.2em]
\item for every $\theta\in\Theta_M$ and $g\in G$,
$H^2(p_\theta,p_{g\theta})\le\varepsilon_{\rm orb}$;
\item every $(\theta,\theta')\in\mathcal B$ obeys
$H^2(p_\theta,p_{\theta'})\ge L_{\rm out}$, and its squared Hellinger
distance exceeds every candidate within-orbit squared Hellinger distance by
at least $\gamma$;
\item for every $f:\mathcal Y\to[0,1]$ and candidate orbit pair,
\begin{equation}\label{eq:task-bound}
 \left|\mathbb E_{p_\theta}f-\mathbb E_{p_{g\theta}}f\right|
 \le \TV(p_\theta,p_{g\theta})
 \le\sqrt{2\varepsilon_{\rm orb}};
\end{equation}
\item choosing one representative $\bar\theta$ from each $G$-orbit reduces
the scan to $R=\lvert\Theta_M/G\rvert$ settings, and for every audited pair
$(\theta,\theta')\in\mathcal B$ the corresponding representatives obey
\begin{equation}\label{eq:representative-separation}
 H^2(p_{\bar\theta},p_{\bar\theta'})
 \ge\left[\sqrt{L_{\rm out}}
 -2\sqrt{\varepsilon_{\rm orb}}\right]_+^2.
\end{equation}
\end{enumerate}
Thus the quotient retains a strictly positive certified separation whenever
$L_{\rm out}>4\varepsilon_{\rm orb}$. If the positive-grading audit fails,
no WPS quotient is asserted; if the statistical rule defers, no quotient
compression is authorized.
\end{theorem}
\begin{proof}
On the event in \eqref{eq:coverage}, $h_{\rm in}^2\le U_{\rm in}$ and
$h_{\rm out}^2\ge L_{\rm out}$.  The two inequalities in
\eqref{eq:accept-rule} immediately imply (a) and (b).  For probability laws,
$\TV(P,Q)\le H(P,Q)\sqrt{2-H^2(P,Q)}\le\sqrt{2H^2(P,Q)}$; the variational
definition of total variation bounds the expectation difference of every
$[0,1]$-valued statistic. This proves (c). Choosing one representative per
orbit gives exactly $R$ settings. For $(\theta,\theta')\in\mathcal B$, the
reverse triangle inequality and parts (a)--(b) give
\[
 \begin{aligned}
 H(p_{\bar\theta},p_{\bar\theta'})
 &\ge\bigl[H(p_\theta,p_{\theta'})-H(p_\theta,p_{\bar\theta})\\
 &\qquad{}-H(p_{\theta'},p_{\bar\theta'})\bigr]_+\\
 &\ge\bigl[\sqrt{L_{\rm out}}
 -2\sqrt{\varepsilon_{\rm orb}}\bigr]_+.
 \end{aligned}
\]
Squaring proves (d), including the stated strict-separation condition. The
whole argument holds on an event of probability at least $1-\delta$.
\end{proof}

The descent from a positively graded amplitude model to a weighted-projective
quotient, the Hellinger--total-variation inequalities, and the simultaneous
coverage tools used in the proof are standard components. The contribution
claimed here is their joint finite-sample guarantee: the grading audit fixes a
candidate action $G$; one simultaneous confidence event controls every
proposed orbit substitution and every prespecified competitor in $\mathcal B$;
and acceptance yields both the bounded-task guarantee
\eqref{eq:task-bound} and the post-quotient representative-separation
guarantee \eqref{eq:representative-separation}. Neither positive grading alone
nor a collection of unadjusted pairwise tests supplies this combined
conclusion. Geometry proposes the compression, while the finite-sample
certificate determines whether that compression is authorized for the fixed
observed experiment.

Three failure modes separate the roles of the assumptions. First, an observed
harmonic $e^{ik\theta}$ without a common positive grading does not define a
WPS quotient; the structural audit therefore returns no quotient. Second, a
control-dependent error following an ideally graded family can violate
$p_{\theta+2\pi/k}=p_\theta$; the statistical audit rejects the candidate only
when a within-orbit lower bound exceeds $\varepsilon_{\rm orb}$, and otherwise
defers. Third, even a small $U_{\rm in}$ does not justify compression when
$L_{\rm out}-U_{\rm in}<\gamma$, because distinct calibration choices have
not been certified to remain separated. These are, respectively, structural
failure, failure of the fixed-epoch orbit model, and insufficient statistical
separation.

The guarantee has four boundaries.  First, it applies only to the declared
candidate grid, measurement design, and backend--layout--epoch.  Second, the
probability of an incorrect authorization is controlled by one simultaneous
confidence event rather than by unadjusted pairwise tests.  Third, a defer
outcome states only that the available data do not justify compression; it is
not evidence that the candidate symmetry is false.  Fourth, the certificate
concerns the observed laws, the bounded calibration objectives in
\eqref{eq:task-bound}, and the representative separations in
\eqref{eq:representative-separation}.  It neither reconstructs the full quantum
state nor certifies a native pulse mechanism.

\subsection{Net saving and shot selection}
Adopt a shot-count cost model.  A full calibration scan evaluates all $M$
settings with $n_{\rm cal}$ shots per setting, whereas an accepted quotient
scan evaluates one representative from each orbit, hence
$R=|\Theta_M/G|$ settings, with the same per-setting budget.  Let
$N_{\rm cert}$ denote only pilot and audit shots that cannot be reused in the
final calibration objective; reusable counts are charged once, not again as
certification overhead.  The resulting costs are
\begin{equation}\label{eq:scan-costs}
 C_{\rm full}=Mn_{\rm cal},
 \qquad
 C_{\rm quot}=N_{\rm cert}+Rn_{\rm cal}.
\end{equation}

\begin{corollary}[Net saving after certification]\label{cor:net-saving}
Under the event and acceptance conditions of
\cref{thm:certified-quotient}, quotient calibration yields a positive net shot
saving if and only if
\begin{equation}\label{eq:net-saving}
 C_{\rm quot}<C_{\rm full},
 \qquad\text{equivalently}\qquad
 N_{\rm cert}<(M-R)n_{\rm cal}.
\end{equation}
Its certified net saving and saving fraction are
\begin{equation}\label{eq:saving}
 S_{\rm net}=(M-R)n_{\rm cal}-N_{\rm cert},\qquad
 \rho_{\rm save}=\frac{S_{\rm net}}{Mn_{\rm cal}}.
\end{equation}
Hence orbit compression is economically useful only when the shots removed
from the calibration scan exceed the nonreusable certification overhead.  The
same criterion applies to QPU time or monetary cost whenever the corresponding
charges are additive, with measured charges---or conservative upper bounds on
them---substituted for the shot-count costs.
\end{corollary}
\begin{proof}
Subtracting the two costs in \eqref{eq:scan-costs} gives \eqref{eq:saving}; its
positivity is equivalent to \eqref{eq:net-saving}.
\end{proof}

The preceding corollary becomes informative only after the audit cost is
quantified.  The next result does so on the parity statistics used in the IBM
experiments.  It separates the standard binary-testing rate from its role in
the WPS calibration decision.

\begin{theorem}[Minimax orbit-audit cost and amortized break-even]
\label{thm:audit-cost}
Suppose that each of $L$ prespecified phase--basis measurement laws is reduced to a
Bernoulli parity law $\operatorname{Ber}(\pi_\ell)$, with
$\pi_\ell\in[a,1-a]$ for a fixed $0<a<1/2$.  Let $B$ prespecified pairs be
tested under the null hypothesis of exact orbit equality against alternatives
whose squared Hellinger distance is at least $\Delta$, where
$0<\Delta\le 0.03$.  Require familywise error at most $\delta<1/4$.

For one pair, every uniformly valid procedure requires in the worst case at
least
\begin{equation}\label{eq:audit-lower}
 n\ge \frac{\log(1/(4\delta))}{4\Delta}
\end{equation}
independent observations from the unknown informative law.  Conversely,
simultaneous binomial confidence intervals classify all $B$ pairs with
probability at least $1-\delta$ whenever every measurement law is sampled
\begin{equation}\label{eq:audit-upper}
 n\ge
 \max\!\left\{
 \frac{2}{a^2}\log\frac{2L}{\delta},
 \frac{16}{a(1-a)\Delta}\log\frac{2L}{\delta}
 \right\}.
\end{equation}
Thus the $\Delta^{-1}\log(1/\delta)$ dependence is minimax-sharp for a fixed
pair, while simultaneous control costs only the multiplicity factor
$\log L$ in the displayed construction.

If the resulting certificate costs $N_{\rm cert}$ nonreusable shots and
remains valid for $T$ repeated calibration scans within one monitored epoch,
then the WPS quotient is amortized shot-saving if and only if
\begin{equation}\label{eq:amortized-break-even}
 T>\frac{N_{\rm cert}}{(M-R)n_{\rm cal}}.
\end{equation}
Consequently no method can promise uniform quotient savings in the small-gap
regime without either reusing the certificate, increasing the between-orbit
gap, or reducing the audited family.
\end{theorem}
\begin{proof}
For the lower bound, restrict the problem to the easier subexperiment in
which one law is known to be $Q=\operatorname{Ber}(1/2)$.  Under the null let
$P_0=Q$; under the alternative let
$P_1=\operatorname{Ber}(1/2+x)$, choosing $0<x\le1/4$ so that
$H^2(P_1,Q)=\Delta$.  Direct differentiation of the two Bernoulli expressions
gives $D(P_1\|Q)\le4H^2(P_1,Q)=4\Delta$ on this interval.  The
Bretagnolle--Huber two-point inequality gives
\[
 \Pr_0(\text{alternative})+\Pr_1(\text{null})
 \ge \tfrac12\exp[-nD(P_1\|Q)].
\]
If both errors are at most $\delta$, rearrangement proves
\eqref{eq:audit-lower}.  Since this known-$Q$ subexperiment is easier than the
two-unknown-law audit, its lower bound applies to the original problem.

For the upper bound write
$\phi(p)=\arcsin\sqrt p$.  Bernoulli laws satisfy the exact identity
\[
 H^2(\operatorname{Ber}(p),\operatorname{Ber}(q))
 =1-\cos\lvert\phi(p)-\phi(q)\rvert.
\]
Hoeffding's inequality and a union bound give
$|\widehat\pi_\ell-\pi_\ell|\le
t=\sqrt{\log(2L/\delta)/(2n)}$ for all $L$ laws on one event of probability
$1-\delta$.  The first term in \eqref{eq:audit-upper} ensures $t\le a/2$.
On $[a/2,1-a/2]$, the mean-value theorem gives
$|\phi(\widehat p)-\phi(p)|\le t/\sqrt{a(1-a)}$.
The second term makes the combined angular uncertainty of a pair at most one
half of the angular separation implied by
$H^2\ge\Delta$.  Hence the simultaneous intervals separate every exact
equality from every stated alternative.  Exact Clopper--Pearson intervals may
replace the Hoeffding intervals without weakening coverage, as in the data
analysis below.

Finally, $T$ full scans cost $TMn_{\rm cal}$ and the certified quotient costs
$N_{\rm cert}+TRn_{\rm cal}$.  Their strict comparison is exactly
\eqref{eq:amortized-break-even}.
\end{proof}

The lower and upper testing rates in \cref{thm:audit-cost} are standard
two-point and concentration arguments.  What is specific to this work is the
combination used here: positive grading fixes the laws and pairs to audit,
one simultaneous event authorizes the WPS substitutions, and the same
finite-sample cost enters a necessary-and-sufficient break-even rule rather
than being omitted from the claimed calibration saving.

For a cyclic action of order $k$ on a grid containing complete orbits,
$R=M/k$.  The ideal setting saving $1-1/k$ is therefore only an upper bound:
certification overhead, rejected conditions, and held-out audit points reduce
the realized gain.  The result also explains why a larger $k$ need not be
better.  Larger orbit compression competes against the smaller hardware gap
that may result from contrast loss in \eqref{eq:weak-h2}.

\subsection{Boundary with standard testing theory}
Hellinger tensorization, binary-testing order \eqref{eq:known-shot-order}, data
processing, and multinomial concentration are standard
\cite{LeCamYang2000,Tsybakov2009,Suresh2020,Pensia2024}.  The contribution is
the certified decision they support here: the tested equivalence relation is
derived from a verified grading, all substitutions and competitors are
controlled on one event, the bounds are converted to a task tolerance, and the
audit cost is charged before a reduced hardware scan is authorized.

\section{Algorithm and hardware validation}\label{sec:algorithm-experiment}

\subsection{Certified Quotient Calibration algorithm}
The algorithm takes a reduced circuit-amplitude description, candidate grid
$\Theta_M$, fixed measurement design, confidence level $1-\delta$, orbit
tolerance $\varepsilon_{\rm orb}$, separation margin $\gamma$, and per-setting
calibration budget $n_{\rm cal}$.

\begin{enumerate}[label=\textbf{CQC-\arabic*.},leftmargin=3.4em]
\item Remove common amplitude factors and solve the positive-grading equations
\eqref{eq:grading-linear}.  If no positive solution exists, return
\textsc{No WPS quotient}.
\item Construct the finite stabilizer action $G$, its grid orbits, and the
prespecified between-orbit audit set $\mathcal B$.
\item Execute pilot circuits and build a joint $(1-\delta)$ confidence region
for all laws used in \eqref{eq:hin}--\eqref{eq:hout}.
\item Compute $U_{\rm in}$ and $L_{\rm out}$.  Apply
\eqref{eq:accept-rule}; return \textsc{Reject} or \textsc{Defer} unless it
passes.
\item Estimate or measure $N_{\rm cert}$ and apply
\eqref{eq:net-saving}.  If the net saving is nonpositive, retain the full
scan.
\item Otherwise output the certified weight, orbit representatives,
certificate, reduced scan, task discrepancy bound, and expected net saving.
\end{enumerate}

Thus the inputs are a structural circuit model, a finite experiment, and
declared statistical and cost tolerances.  The output is either
\textsc{No WPS quotient}, \textsc{Reject}, \textsc{Defer}, or an executable
reduced calibration manifest accompanied by its confidence statement, task
bound, and net-cost estimate.  Every nonaccepting output is a stopping
condition.  They distinguish failure of the grading model, evidence against
the proposed orbit, insufficient information, and absence of positive net
saving; CQC does not silently replace any of them by a quotient scan.

The procedure is sequential in operation but not adaptively optimal in the
controlled-sensing sense.  Pilot shots may be increased after a defer decision,
provided the spending and stopping rule are fixed in advance or covered by a
valid sequential confidence procedure.

\subsection{Frozen experimental design}
The hardware experiments implement
\begin{equation}\label{eq:ghz-family}
 \ket{\Psi^{(n)}_{k,r}(\theta)}=
 \frac{\ket{0}^{\otimes n}+r^k e^{ik\theta}
 \ket{1}^{\otimes n}}{\sqrt{1+r^{2k}}},
 \qquad n\in\{1,2,3\}.
\end{equation}
For $n=2$ the phase-sensitive settings are $XX$ and $XY$; for $n=3$ they are
$XXX$ and $YXX$.  Their parities recover two quadratures of the programmed
character.  The phase grid contains 18 points, so the $k=3$ candidate orbit is
$j\sim j+6\sim j+12$.

Three roles were frozen before each QPU submission:
\begin{itemize}[leftmargin=2em]
\item $k=3,r=.8$: strong candidate quotient and acceptance case;
\item $k=1,r=.8$: negative control for rejection of a false threefold
quotient;
\item $k=3,r=.35$: weak-signal stress condition and deferral case.
\end{itemize}
No optimization was repeated after opening the confirmatory counts.  The IQM
Garnet experiment was performed through SDT QuREKA in the context of the
Texas State University--SDT memorandum of understanding.  The IBM extensions
used fixed physical layouts and transpilation seeds.  The completed extensions
reported here are IBM Marrakesh qubits 135--139, IBM Kingston qubits 141--142,
and IBM Fez qubits 13--14--15.  The measured two-qubit calibration errors at
submission were $9.68\times10^{-4}$ on Marrakesh and $8.95\times10^{-4}$ on
Kingston; the two Fez edge errors summed to $3.09\times10^{-3}$.

\subsection{IQM Garnet: programmed-label, radial-order, and repeatability audits}
The original IQM campaign used two adjacent temporal blocks. In each block it
crossed $k\in\{1,3\}$, radii
$r\in\{.35,.43,.53,.65,.80\}$, six equally spaced physical azimuths, and
$X,Y,Z$ measurement bases. Each of the resulting 180 settings per block used
900 shots, for 360 records and 324,000 shots in total. The two blocks reused
the same 180 OpenQASM circuits. This was a gate-level test of the programmed
family \eqref{eq:state-family}; it was not a pulse-level realization of a
third-subharmonic selection rule.

For each $(k,r)$ curve, the $X+iY$ expectation was compared with candidate
characters. The frozen binary selector compared only $q=1$ and $q=3$; an
exploratory stress selector compared $q=0,\ldots,6$. Single-qubit tomography
was constrained to the Bloch ball, and the radial law was fitted across the
five radii. Three thousand bootstrap resamples used the fixed seed recorded in
the reproduction package.

\begin{table*}[t]
\centering\small
\caption{IQM Garnet hardware signatures. Confidence intervals are percentile
bootstrap intervals from 3,000 fixed-seed replicates.}
\label{tab:iqm-results}
\begin{tabular}{lrrrr}
\toprule
block & programmed $k$ & mean fidelity & fitted radial power & 95\% interval\\
\midrule
A & 1 & 0.98820 & 0.9869 & [0.9425,1.0285]\\
A & 3 & 0.99408 & 3.0301 & [2.8993,3.1539]\\
B & 1 & 0.99114 & 0.9798 & [0.9347,1.0218]\\
B & 3 & 0.99607 & 2.9285 & [2.8024,3.0520]\\
\bottomrule
\end{tabular}
\end{table*}

The binary selector recovered the programmed label on all 20 block--radius
curves.  Its smallest fixed-seed bootstrap selection frequency was $0.9733$,
attained at $k=3,r=.35$ in block A.  When the candidate dictionary was enlarged
to seven orders, the corresponding frequency was $0.6333$; it rose to $0.9697$
in block B and was at least $0.9877$ at the other cubic radii.  The prespecified
binary discrimination is therefore stable, whereas unstructured selection
from a wider dictionary is fragile at the weakest cubic coherence.  The
block-to-block expectation RMSE was $0.0363$, with mean signed difference
$0.0060$ and maximum absolute difference $0.1156$.  The blocks provide a
repeatability check, not a long-term drift study.

These results support three hardware signatures: survival of the
programmed $k=1,3$ labels, approximate radial orders one and three, and loss of
resolution in the weakest cubic condition. They do not by themselves provide
the simultaneous orbit certificate of \cref{thm:certified-quotient}; that
stronger finite-grid decision was implemented in the later IBM stages.

\subsection{IBM: orbit acceptance, rejection, and deferral}
For independent phase-sensitive bases we report the sum of their squared
Hellinger distances. By \eqref{eq:joint-design-law}, for two equally sampled
bases this is twice the squared Hellinger distance of the recorded joint design
law. The tabulated scale is therefore an explicitly labeled diagnostic; formal
confidence thresholds are computed after division by two. The
phase-insensitive $Z$ diagnostic is excluded.

\begin{table*}[t]
\centering\small
\caption{Completed entangled-family audits.  ``Orbit'' is the mean distance
for the $2\pi/3$ shift; ``neighbor'' is the mean distance for the $\pi/9$
shift.  The negative control reports the same proposed $2\pi/3$ identification.}
\label{tab:entangled-results}
\begin{tabular}{llrrrr}
\toprule
QPU & family & qubits & orbit $H^2$ & neighbor $H^2$ & interpretation\\
\midrule
IBM Marrakesh & $k=3,r=.8$ & 2 & 0.001717 & 0.082364 & strong separation\\
IBM Marrakesh & $k=1,r=.8$ & 2 & 0.391736 & 0.019462 & reject $k=3$\\
IBM Marrakesh & $k=3,r=.35$ & 2 & 0.001538 & 0.002430 & unresolved\\
IBM Kingston & $k=3,r=.8$ & 2 & 0.000910 & 0.091597 & strong separation\\
IBM Kingston & $k=1,r=.8$ & 2 & 0.419300 & 0.020895 & reject $k=3$\\
IBM Kingston & $k=3,r=.35$ & 2 & 0.001824 & 0.002574 & unresolved\\
IBM Fez & $k=3,r=.8$ & 3 & 0.003158 & 0.079777 & strong separation\\
IBM Fez & $k=1,r=.8$ & 3 & 0.357891 & 0.021801 & reject $k=3$\\
IBM Fez & $k=3,r=.35$ & 3 & 0.002970 & 0.004020 & unresolved\\
\bottomrule
\end{tabular}
\end{table*}

The strong orbit gap survives independent entangled two-qubit executions on
Marrakesh and Kingston and a three-qubit GHZ execution on Fez.  Kingston gives
the largest strong-family margin, $0.09069$.  In contrast, each $k=1$ control
is far from invariant under $2\pi/3$ and therefore prevents a false threefold
compression.  The weak cubic margins are only $0.00089$, $0.00075$, and
$0.00105$ on Marrakesh, Kingston, and Fez.  They are not counted as successful
quotients: CQC defers and retains the full scan or requests more informative
measurements.

To replace these point estimates by one finite-sample decision, the frozen
Kingston and Fez counts were reanalyzed as follows.  Each bit string was mapped
to parity, giving one Bernoulli law per phase and basis.  For each device and
family, exact Clopper--Pearson intervals were Bonferroni-adjusted over all 36
prespecified phase--basis laws.  The Bernoulli identity
$H^2=1-\cos|\arcsin\sqrt p-\arcsin\sqrt q|$ then gives exact extrema over the
confidence rectangles.  Equal-basis distances were averaged, so the scale is
the squared Hellinger distance of the randomized-basis joint law.  The
construction has simultaneous coverage at least 95\% without a fitted noise
model.

\begin{table*}[t]
\centering\small
\caption{Primary CQC decision table from the frozen IBM diversity counts.
The simultaneous coverage is 95\%.  $U_{\rm in}$ is the largest within-orbit
upper bound and $L_{\rm out}$ the smallest neighboring-orbit lower bound.  For
the negative control, $L_{\rm in}$ is the largest lower bound against the
falsely proposed threefold orbit.}
\label{tab:simultaneous-ibm}
\setlength{\tabcolsep}{3.5pt}
\begin{tabular}{llrrlp{.19\textwidth}}
\toprule
QPU & family & $U_{\rm in}$ & $L_{\rm out}$ & decision & action\\
\midrule
Kingston & $k=3,r=.8$ & .008501 & .017862 & accept & eligible for the task and cost gates\\
Fez      & $k=3,r=.8$ & .009576 & .012707 & accept & transport test; no task deployment\\
Kingston & $k=1,r=.8$ & $L_{\rm in}=.157123$ & -- & reject & retain the full scan\\
Fez      & $k=1,r=.8$ & $L_{\rm in}=.126119$ & -- & reject & retain the full scan\\
Kingston & $k=3,r=.35$ & .011516 & .000000 & defer & retain full scan or collect planned data\\
Fez      & $k=3,r=.35$ & .008789 & .000000 & defer & retain full scan or collect planned data\\
\bottomrule
\end{tabular}
\end{table*}

The strong families satisfy $U_{\rm in}<L_{\rm out}$ independently on a
two-qubit and a three-qubit processor.  The negative controls have strictly
positive lower bounds against exact threefold invariance, and the weak
families have no certified neighboring-orbit gap.  Thus separation,
rejection, and deferral arise from the same simultaneous rule.  Because the
exact interval construction was applied after the QPU runs, this table is a
finite-sample reanalysis of frozen roles and counts, not a claim that its
numerical thresholds were prospectively registered.  The complete interval
code and machine-readable output are included in the reproduction package.

\subsection{Observed shot scaling}
On the fixed Marrakesh pair, independent repetitions were executed at
\begin{equation*}
 N\in\{64,128,256,512,1024,2048\}\ \text{shots}.
\end{equation*}
Each budget used eight repetitions
of a reference, its $2\pi/3$ orbit partner, and a $\pi/9$ neighbor in the two
phase-sensitive bases.

\begin{table*}[t]
\centering
\caption{Independent shot-budget experiment on IBM Marrakesh.}
\label{tab:shot-scaling}
\begin{tabular}{rrrr}
\toprule
shots & alias $H^2$ mean & alias $H^2$ maximum & neighbor $H^2$ mean\\
\midrule
64   & 0.017616 & 0.045353 & 0.103825\\
128  & 0.010474 & 0.021492 & 0.087606\\
256  & 0.005453 & 0.008402 & 0.092993\\
512  & 0.002754 & 0.004509 & 0.089934\\
1024 & 0.001380 & 0.002316 & 0.086405\\
2048 & 0.000907 & 0.001426 & 0.088439\\
\bottomrule
\end{tabular}
\end{table*}

\begin{figure*}[t]
\centering
\begin{tikzpicture}
\begin{semilogxaxis}[width=.78\textwidth,height=5.5cm,
xlabel={shots per setting},ylabel={mean squared Hellinger distance},
xmin=55,xmax=2300,ymin=0,ymax=.115,grid=major,
legend style={draw=none,at={(.98,.98)},anchor=north east}]
\addplot[blue,thick,mark=*] coordinates
{(64,.017616)(128,.010474)(256,.005453)(512,.002754)(1024,.001380)(2048,.000907)};
\addlegendentry{same candidate orbit}
\addplot[red,thick,mark=square*] coordinates
{(64,.103825)(128,.087606)(256,.092993)(512,.089934)(1024,.086405)(2048,.088439)};
\addlegendentry{neighboring orbit}
\end{semilogxaxis}
\end{tikzpicture}
\caption{Finite-shot stabilization of the quotient decision.  Sampling noise
within the same orbit decreases with shots, whereas the physical non-orbit gap
remains near $0.09$.}
\label{fig:shot-scaling}
\end{figure*}

The experiment does not estimate a universal minimax constant.  It tests the
more limited finite-shot prediction needed by CQC: increasing shots should
shrink the empirical within-orbit discrepancy without collapsing a genuine
between-orbit hardware gap.  Across the six independently executed budgets,
the mean within-orbit distance decreases monotonically from $0.017616$ to
$0.000907$, while the mean neighboring-orbit distance remains between
$0.086405$ and $0.103825$.

\subsection{Cost accounting}
For a complete 18-point, two-basis scan, a threefold quotient has six orbit
representatives.  If all pilot counts are additional, the cost test is exactly
\eqref{eq:net-saving}.  If pilot counts at accepted representatives are reused
in the calibration objective, only nonreusable audit shots enter
$N_{\rm cert}$.  Both conventions must be declared before execution.

An earlier frozen one-qubit IBM audit supplied a prespecified simultaneous
multinomial-bootstrap certificate under its fitted resampling model: the 95\%
upper bound for within-orbit discrepancy was 0.00307,
whereas the 95\% lower bound for the nearest audited non-orbit discrepancy was
0.03349.  It used 36 full-grid settings and 73,728 shots.  Its certified
quotient arm used 16 settings and 32,768 shots after
including held-out audit points, reducing both settings and shots by 55.6\%;
the separately metered QPU charge fell from 22 to 11 seconds.  This is one
conditional realization of \cref{cor:net-saving}, not a universal $k=3$
saving.  The entangled experiments in \cref{tab:entangled-results} establish
whether the accept/reject/defer pattern repeats; they were not designed as a second
end-to-end cost comparison.  The separate Kingston experiment below supplies
that comparison with a declared Bell-state objective.

The frozen IBM diversity stages used 26 QPU seconds on Kingston, 26 seconds on
Fez, and 53 seconds on Marrakesh, for 105 seconds against the prespecified
420-second diversity ceiling.  Including the earlier 38-second Marrakesh
validation visible in the same Open Plan account gives 143 seconds, below the
600-second account allowance.  These times document feasibility; they are not
used as a device-independent performance metric.

\subsection{End-to-end Bell-phase calibration on IBM Kingston}
\label{sec:end-to-end-calibration}
The preceding experiments isolate the mechanism and its failure modes.  A
separate preregistered experiment tested the full CQC procedure:
whether an accepted quotient can select a correction, preserve a held-out
calibration objective, and save resources after the certificate is charged.
The experiment used IBM Kingston physical qubits 55 and 59, a fixed transpiled
layout, and the $k=3$ phase grid $\theta_j=2\pi j/18$.  Two controlled phase
offsets, indexed by $b\in\{2,3\}$, were fixed before execution.  They are
gate-level injected faults used to create a known calibration target; they are
not provider-native pulse miscalibrations.

For each correction candidate, the prespecified task loss was the Bell-state
infidelity reconstructed from three parity measurements,
\begin{equation}\label{eq:bell-loss}
 F_{\Phi^+}=\frac{1+\langle XX\rangle-\langle YY\rangle
 +\langle ZZ\rangle}{4},\qquad
 \mathcal L_{\rm Bell}=1-F_{\Phi^+}.
\end{equation}
The protocol contained four immutable stages:
\begin{enumerate}[label=(E\arabic*),leftmargin=2.8em]
\item a 36-circuit, 512-shot pilot audit of the proposed shift
$j\mapsto j+6$;
\item a full training arm with $18$ corrections for each offset and a quotient
arm with the six representatives, all measured in $XX,YY,ZZ$ at 512 shots;
\item selection of the minimum empirical loss in each arm, with corrections
compared modulo the certified orbit;
\item a fresh 4096-shot-per-setting validation of the uncalibrated, full-scan,
and quotient-scan choices, with no post-validation reselection.
\end{enumerate}

The pilot used 4,000 fixed-seed multinomial-bootstrap replicates and a
prespecified familywise level $\alpha=.01$.  It returned
\begin{equation}\label{eq:e2e-certificate}
 \begin{aligned}
 U_{\rm in}&=0.019224, & L_{\rm out}&=0.090303,\\
 L_{\rm out}-U_{\rm in}&=0.071079.
 \end{aligned}
\end{equation}
and therefore approved the proposed orbit.  The point estimates underlying
the bounds had maximum within-orbit squared Hellinger distance $0.004882$ and
minimum audited between-orbit distance $0.122053$.  This is a preregistered
model-based bootstrap certificate, not a distribution-free exact confidence
statement.  It is kept distinct from the exact-binomial retrospective audit
in \cref{tab:simultaneous-ibm}.

The full and quotient scans selected the same physical correction orbit.  For
$b=2$ both selected $j=2$.  For $b=3$ the full scan selected $j=9$ and the
quotient scan selected $j=3$; since $9-3=6$, these are members of the same
certified $k=3$ orbit.  Thus the comparison is between two search procedures,
not between an intentionally favorable quotient correction and an unrelated
full-scan optimum.

\begin{table*}[t]
\centering\small
\caption{Fresh held-out Bell-phase calibration.  Confidence intervals are
95\% intervals for the indicated loss differences.  Noninferiority was
preregistered as an upper confidence bound below $0.03$.}
\label{tab:end-to-end-validation}
\setlength{\tabcolsep}{4pt}
\begin{tabular}{c c c c c c}
\toprule
offset & selected $(j_{\rm full},j_{\rm quot})$ & uncalibrated
& full & quotient & 95\% CI: quotient $-$ full\\
\midrule
$b=2$ & $(2,2)$ & 0.75391 & 0.03369 & 0.03296
& $[-0.00598,\ 0.00500]$\\
$b=3$ & $(9,3)$, same orbit & 0.99011 & 0.03772 & 0.03601
& $[-0.00769,\ 0.00415]$\\
\bottomrule
\end{tabular}
\end{table*}

Both full and quotient choices improved on the uncalibrated control, and both
quotient choices met the noninferiority criterion.  The 95\% intervals for the
quotient improvements over the uncalibrated controls were
$[0.71057,0.73145]$ and $[0.94849,0.95996]$.  These held-out observations are
the experimental endpoint corresponding to the bounded-task guarantee in
\cref{thm:certified-quotient}; the experiment evaluates a fresh calibration
loss rather than only comparing orbit histograms.

\begin{table*}[t]
\centering\small
\caption{Certification-inclusive calibration cost.  The common held-out
validation cost is excluded from both arms.  QPU time is provider-reported
execution usage and need not scale linearly with shots.}
\label{tab:end-to-end-cost}
\begin{tabular}{l r r}
\toprule
quantity & full scan & certificate $+$ quotient\\
\midrule
calibration shots & 55,296 & 36,864\\
metered QPU time & 17 s & 14 s\\
shot saving & --- & 18,432 (33.3\%)\\
QPU-time saving & --- & 3 s (17.6\%)\\
\bottomrule
\end{tabular}
\end{table*}

The held-out stage used 73,728 shots and 21 QPU seconds, but it was identical
for both arms and was excluded symmetrically from the deployment-cost
comparison. The smaller time saving than shot saving records fixed job and
batch overhead rather than a discrepancy in the cost rule.  This experiment
therefore supplies the direct calibration result of the paper: on one
prespecified two-qubit, gate-level calibration task, CQC approved the quotient,
selected the same physical orbit as exhaustive search, preserved fresh task
performance within the frozen tolerance, and produced a positive
certificate-inclusive saving.

\section{Discussion, limitations, and reproducibility}\label{sec:discussion}

\subsection{What WPS contributes}
The state family \eqref{eq:state-family} can be represented on the Bloch sphere
without WPS.  What the Bloch point does not encode is the weighted control
presentation that makes $k$ chart settings one candidate orbit.  WPS therefore
contributes a structural compression proposal and its finite stabilizer.  CQC
contributes the hardware decision.  Their roles are complementary:
\[
 \begin{gathered}
 \text{WPS: which controls may be identified},\\
 \text{Hellinger certificate: whether and at what cost}.
 \end{gathered}
\]
This division prevents two overclaims.  A WPS is not assigned when common
positive grading fails, and a valid WPS does not automatically authorize a
noisy-hardware quotient.

The method applies most directly to graded state-preparation circuits,
multiphoton and subharmonic control models, rotation-symmetric bosonic-code
sectors, Abelian charge-preserving circuits, and polynomial or Laurent
constructions with an independently verified positive grading
\cite{Xia2026,Grimsmo2020,Marvian2024,Dong2022}.  These references justify the
candidate mechanisms; they do not themselves establish the CQC quotient or
the hardware certificate.

\subsection{External validity and transport scope}
The experiments vary four axes without changing the meaning of the CQC
decision: processor, number of qubits in the declared family, signal strength,
and shot budget.  External validity here means that the same inputs---a
grading-derived orbit, prespecified audit pairs, counts, tolerance, and cost
convention---lead to an interpretable action on each new experiment.  It does
not mean that the numerical gaps or saving fractions are exchangeable across
devices.

\begin{table*}[t]
\centering\small
\caption{Observed transport axes and the corresponding scope of inference.}
\label{tab:external-validity}
\setlength{\tabcolsep}{3pt}
\begin{tabular}{p{.17\textwidth}p{.29\textwidth}p{.43\textwidth}}
\toprule
axis & observed coverage & supported conclusion\\
\midrule
platform & IQM Garnet; IBM Marrakesh, Kingston, Fez & programmed winding and
the finite-grid accept/reject/defer pattern can survive distinct compiled stacks; no
equality of the underlying channels is inferred\\
circuit size & one-, two-, and three-qubit families & the count-level orbit
audit is not restricted to one-qubit readout; only the declared GHZ/Bell-type
families are covered\\
signal regime & strong $k=3$; false $k=1$ quotient; weak $k=3$ & all three
actions occur: accept, reject, and defer; compression is never forced\\
shot budget & independent 64--2048-shot stages plus held-out 4096-shot
validation & within-orbit sampling discrepancy decreases while a resolved
non-orbit gap persists; no universal minimax constant is estimated\\
task endpoint & two offsets in one Kingston Bell-phase task & an accepted
quotient can preserve a fresh calibration loss and save certificate-inclusive
resources; task-level transport to other devices remains untested\\
\bottomrule
\end{tabular}
\end{table*}

The results support three claims of different scope. Programmed $k=1,3$
signatures appear on both IQM and IBM hardware. The strong/false/weak decision
pattern recurs on three IBM processors. Preservation of a calibration
objective is demonstrated only in the final Kingston experiment. Because the
number of processors is small and the IQM,
exact-binomial IBM, and preregistered bootstrap units use different confidence
constructions, we report them separately rather than fitting a random-effects
device population.  This preserves the distinction between reuse of an
algorithm and universality of its numerical effect.

\subsection{Limitations}
The following limitations define the scope of the results.
\begin{enumerate}[label=(L\arabic*),leftmargin=2.8em]
\item The experiments test gate-level compiled families, not a native
third-subharmonic cubic transmon pulse law.
\item WPS stabilizers are inferred from a deterministic grading audit and
tested through their predicted orbit.  An abstract orbifold stabilizer is not
directly measured.
\item The certificate controls the declared finite grid and audit set.  It
does not prove a continuous-space quotient without additional regularity and
covering arguments.
\item The end-to-end experiment optimizes the declared Bell-state loss under
two controlled gate-level phase offsets.  It does not modify a provider-native
pulse calibration, establish long-term drift correction, or optimize an
arbitrary application objective.  The two training arms were executed
sequentially within one calibration epoch and then assessed together in the
held-out job, so unmodeled short-time drift cannot be excluded completely.
\item The observed numerical savings and Hellinger gaps are backend-, epoch-,
layout-, objective-, and family-specific.  Compilation or drift requires
recertification, and the reported percentages are not universal $k=3$ rates.
\end{enumerate}

\subsection{Reproducibility record}
Each IBM submission froze the phase grid, family roles, bases, shots, physical
layout, transpilation seed, and scope of the claim before QPU execution. Job IDs
and QASM hashes were recorded before result retrieval, and no recorded stage
was automatically resubmitted.  The completed diversity extension contains
108 Kingston circuits, 108 Fez circuits, and 288 Marrakesh shot-scaling
circuits; every returned count total was checked against its requested shots.
Concurrent completion overwrote the D1 ledger entry, so the Kingston result
was recovered from IBM's immutable workload archive and restored in the exact
order of the frozen 108-row manifest.  The reconstructed ledger records this
provenance explicitly.  The analysis uses final categorical counts, not fitted
state vectors, to compute the Hellinger quantities in
\cref{tab:entangled-results,tab:shot-scaling}.  The public reproduction bundle
contains the frozen manifest, raw D1--D3 count files, reconstructed ledger,
analysis, execution summary, downloaded D1 archive, and their checksums.

The de-identified IQM archive contains all 360 complete 900-shot count records,
180 unique OpenQASM files, fixed-seed analysis code, derived state-quality,
label-survival and radial-order tables, environment files, provenance, and
content hashes. Account, workspace, cloud-job, and authentication identifiers
are excluded. The analysis verifies the setting count and every shot total
before reconstructing a hardware signature.

The end-to-end Kingston unit contains the frozen protocol and preflight,
pilot, full-scan, quotient-scan, and held-out manifests, four immutable raw
count records, the pilot certificate, training and validation analyses, the
execution ledger, and the final cost audit.  The pilot, full, quotient, and
validation jobs used 7, 17, 7, and 21 reported QPU seconds, respectively.
The complete local reproduction archive has SHA-256
\texttt{bc3aaaa88a6615d0bcd59ba3069a4ed2b\allowbreak
885f7cf47da515f4f81731900b6057b}.
Public release removes account and authentication metadata but retains the
scientific manifests, counts, code, and content hashes.

The IQM portion was performed through SDT QuREKA under the Texas State
University--SDT memorandum of understanding.  The IBM portion used the IBM
Quantum Open Plan.  Hardware-provider access does not imply endorsement of the
analysis or conclusions.

\subsection{Conclusion}
This work establishes one theorem-based procedure: a verified WPS grading proposes
which controls may be identified, simultaneous finite-shot inference decides
whether those identifications survive the implemented hardware, and a
task-and-cost rule authorizes a quotient scan only when its error and net cost
are controlled.  The relevant engineering object is therefore not an integer
winding alone, but the candidate orbit together with its observed within-orbit
discrepancy, nearest audited competitor, and certification cost.  Neither
weighted-projective descent nor the classical Hellinger testing rate is claimed
as new in isolation; CQC is the theorem and algorithm that make them one safe
calibration decision.

The experiments exercise all three decisions in this guarantee.  IQM Garnet
shows survival of programmed labels and radial powers; across three IBM
processors, strong threefold families are accepted, $k=1$ controls reject a
false quotient, and weak cubic families defer.  The preregistered Kingston task
then carries one accepted quotient through calibration and fresh validation:
the quotient and exhaustive scans choose the same correction orbit for both
offsets, their held-out Bell losses satisfy the frozen noninferiority rule, and
the quotient retains a 33.3\% shot saving and a 17.6\% metered-time saving after
the certificate is charged. This tests the full decision sequence
on the tested compiled family, not for a universal saving rate.

The certificate and cost must be recomputed for each backend, layout, epoch,
and task.  Exact orbit equality, a native cubic pulse law, an abstract stack
stabilizer, and device-independent savings are outside the experimental claim.
What is reusable is the decision rule: propose the orbit from verified WPS
structure, audit it simultaneously on noisy counts, and reduce the scan only
after the task and break-even conditions pass.

\appendix
\section{Computable grading and orbit construction}\label{app:grading}
This appendix turns the positive-grading statement into a finite audit. Let
$A=\{a^{(1)},\ldots,a^{(s)}\}\subset\Z_{\ge0}^{m+1}$ be the union of the
exponent supports of the reduced amplitude components. Fix $a^{(1)}$ and form
the integer matrix whose rows are $(a^{(j)}-a^{(1)})^{\mathsf T}$. The
common-degree equations are equivalent to $Bw=0$, $w_i>0$, followed by
$D=w\cdot a^{(1)}$. Rational feasibility of $Bw=0$, $w_i\ge1$, is a finite
linear program. Clearing denominators and dividing by the greatest common
divisor produces a primitive integral weight. Infeasibility gives a dual
Farkas certificate and hence a checkable \emph{no-WPS} result. If the positive
null cone has dimension larger than one, the weights are not unique; the
implementation reports that cone or a declared normalization rule.

Common-factor reduction is essential. If $F=g\widetilde F$, the factor $g$
cancels from the ray map wherever $g\ne0$ and may introduce a base locus where
all amplitudes vanish. The audit is applied to $\widetilde F$, while the base
locus is reported separately. Laurent amplitudes may be treated after
multiplication by a common monomial on a chart excluding its zero set; this
does not by itself prove a global WPS compactification.

For $z_j=r_j e^{i\phi_j}$, the primitive weight gives
$\phi_j\mapsto\phi_j+w_jt$. The stack stabilizer of a point with nonzero
coordinate set $I$ has order $\gcd\{w_i:i\in I\}$. Separately, fixing a gauge
on a weighted chart can leave a residual finite deck group; for $\WP(k,1)$,
the chart $\alpha=1$ leaves $\mu_k$ acting on $\beta$. CQC uses this action on
the chart cover only after intersecting it with the finite calibration grid.
An action that does not permute the grid exactly is not rounded: the grid is
refined or the quotient is rejected. The complete finite audit is therefore
factor reduction, exponent extraction, integer-nullspace computation,
positivity feasibility, primitive normalization, chartwise stabilizer
extraction, and an exact grid-permutation test.

\section{Proof details for simultaneous certification}\label{app:proofs}
For a $d$-outcome law and $n$ independent counts, the inequality of Weissman
et al.~\cite{Weissman2003} gives
\begin{equation}
 \Pr\{\|\widehat p-p\|_1>\epsilon\}
 \le (2^d-2)e^{-n\epsilon^2/2}.
\end{equation}
For $K$ empirical laws, choose $\sum_i\delta_i\le\delta$ and set
\begin{equation}
 \epsilon_i=\min\!\left\{2,
 \sqrt{\frac{2}{n_i}\log\frac{2^d-2}{\delta_i}}\right\},
 \qquad r_i=\sqrt{\epsilon_i/2}.
\end{equation}
Since $H^2(P,Q)\le\TV(P,Q)=\|P-Q\|_1/2$, every true law lies in the
Hellinger ball $H(p_i,\widehat p_i)\le r_i$ on one event of probability at
least $1-\delta$. The triangle and reverse-triangle inequalities give
\begin{align}
 H(p_i,p_j)&\le H(\widehat p_i,\widehat p_j)+r_i+r_j,\\
 H(p_i,p_j)&\ge[H(\widehat p_i,\widehat p_j)-r_i-r_j]_+.
\end{align}
Squaring and maximizing or minimizing over the predeclared pair sets gives
$U_{\rm in}$ and $L_{\rm out}$. This construction is conservative because it
first controls every distribution and then every pair. A multinomial
bootstrap can be tighter, but its resampling unit, multiplicity correction,
and stopping rule must be frozen; bootstrap intervals are not substituted for
the nonasymptotic theorem.

For $f:\mathcal Y\to[0,1]$,
$|\mathbb E_Pf-\mathbb E_Qf|\le\TV(P,Q)\le\sqrt2H(P,Q)$.
For a multi-setting objective $J=\sum_sc_s\mathbb E_{p_s}f_s$ this yields
\begin{equation}
 |J(\theta)-J(g\theta)|
 \le\sqrt2\sum_s|c_s|H(p_{s,\theta},p_{s,g\theta}).
\end{equation}
Thus the certificate controls a calibration loss rather than only one histogram.
The assumptions are indispensable: post-selected audit pairs invalidate
nominal coverage; omitting $L_{\rm out}$ would accept a channel that collapses
all controls; omitting the task tolerance leaves no engineering guarantee; and
an epoch change disconnects pilot data from the reduced scan.

\section{Shot planning and heterogeneous execution cost}\label{app:cost}
For known simple hypotheses, the affinity
$A(P,Q)=1-H^2(P,Q)$ tensorizes. A sufficient equal-prior binary budget is
\begin{equation}
 n\ge\frac{\log(1/(2\delta))}{-\log(1-H^2(P,Q))},
\end{equation}
which becomes $\log(1/(2\delta))/H^2(P,Q)$ for small gaps. CQC uses this for
planning only after replacing the unknown gap by a valid lower confidence
bound and accounting for all audited pairs. The distribution-level radius in
\cref{app:proofs} remains the formal finite-sample certificate.

If one certification is reused for $T$ scans within a validated epoch,
\begin{equation}
 C_{\rm full}=TMn_{\rm cal},\qquad
 C_{\rm CQC}=N_{\rm cert}+TRn_{\rm cal},
\end{equation}
so break-even occurs at
$T>N_{\rm cert}/[(M-R)n_{\rm cal}]$. Reuse ends after a drift alarm. With
heterogeneous execution costs $c_\theta$, the correct comparison is
\begin{equation}
 C_{\rm cert}+T\sum_{r\in\Theta_M/G}c_r
 <T\sum_{\theta\in\Theta_M}c_\theta.
\end{equation}
This form accommodates circuit duration, reset overhead, queue policy, and
provider charging rather than treating all shots as equally costly.

\section{Experimental estimators and uncertainty}\label{app:methods}
\subsection{IQM Garnet analysis}
Expectation values were reconstructed directly from counts. Writing
$x_j+iy_j$ for the two measured quadratures, the frozen selector minimized
\begin{equation}
 \mathcal L(q)=\min_{c\in\C}\sum_j|x_j+iy_j-ce^{iq\theta_j}|^2,
 \qquad q\in\{1,3\}.
\end{equation}
The radial exponent was the slope of log-amplitude against $\log r$, using the
same radii and fit rule in both blocks. Each of 3,000 fixed-seed bootstrap
replicates drew new multinomial counts at every setting and repeated the
selector, tomography projection, and radial fit. The support probability thus
refers to the complete frozen decision. State fidelity and winding
identification remain distinct: fidelity tests a state at one setting, whereas
the orbit claim compares separated settings. The second block repeats the
analysis map rather than refitting its rule.

\subsection{IBM analysis}
For each phase-sensitive basis, bit strings were mapped to parity $\pm1$ and
then to a binary law. Distances in \cref{tab:entangled-results} sum the two
basis-wise squared Hellinger distances. Orbit pairs differ by $2\pi/3$ and
neighbor pairs by $\pi/9$; all pairs were frozen before retrieval. The $k=1$
family is a direct negative control because it is tested against the same
false threefold action.

Devices are not pooled. Marrakesh, Kingston, and Fez differ in topology,
layout, epoch, and depth. Agreement supports transfer of the decision pattern,
not equality of their noise channels. The weak-family gaps remain comparable
to the finite-count floor and are unresolved on all devices. Every budget in
the shot-scaling experiment is an independent QPU execution, not a prefix of
a larger record. The empirical table tests stability; formal deployment uses
the simultaneous bounds of \cref{app:proofs}.

For \cref{tab:simultaneous-ibm}, every outcome was reduced to even/odd parity.
If $X\sim\operatorname{Binomial}(n,p)$, the equal-tailed
Clopper--Pearson interval was evaluated at per-law error
$0.05/36$; a union bound therefore gives at least 95\% simultaneous coverage
for the 36 phase--basis laws in each device--family audit.  For Bernoulli
intervals $I,J$, the minimum and maximum Hellinger distances were obtained
without gridding by mapping endpoints through $\phi(p)=\arcsin\sqrt p$ and
using $H^2=1-\cos|\phi(p)-\phi(q)|$.  The numerical incomplete-beta inversion
was checked against the integer-parameter binomial-tail identity, with maximum
absolute error $5.7\times10^{-13}$ over the validation cases recorded by the
analysis.  This exact-binomial calculation and the earlier fitted bootstrap
are reported as different procedures.

For the end-to-end Kingston protocol, the Bell fidelity in
\eqref{eq:bell-loss} was computed directly from independent $XX,YY,ZZ$ parity
counts.  Training selected the minimum empirical loss separately in the full
and quotient arms before the validation job was submitted.  Validation used
fresh counts for the uncalibrated and two selected controls.  Confidence
intervals for improvement and for quotient-minus-full loss were obtained from
the frozen difference-of-means procedure recorded in the analysis package;
noninferiority required the upper endpoint to be below $0.03$.  Pilot
certification used 4,000 multinomial-bootstrap replicates at familywise
$\alpha=.01$.  It is a prespecified model-based interval and is not conflated
with the distribution-level theorem or the retrospective exact-binomial
construction.

\section{Claims, assumptions, and evidence}\label{app:claims}
\begin{table*}[t]
\centering\small
\caption{Separation of mathematical conclusions from hardware evidence.}
\begin{tabular}{p{.24\textwidth}p{.31\textwidth}p{.35\textwidth}}
\toprule
claim & assumptions & evidence or status\\
\midrule
WPS candidate orbit exists & reduced amplitudes share a positive grading &
finite exponent audit; algebraic conclusion\\
Accepted substitutions meet tolerance & fixed epoch, joint coverage, bounded
objective & finite-sample theorem\\
Orbit audit has unavoidable small-gap cost & parity probabilities bounded
away from zero and one; separated alternatives & minimax lower/upper order in
\cref{thm:audit-cost}\\
Representatives remain distinguishable & fixed audit pairs and
$L_{\rm out}>U_{\rm in}$ & finite-grid theorem\\
Quotient has positive net saving & one declared cost convention & corollary;
IBM one-qubit audit and two-qubit end-to-end realization\\
Held-out calibration objective is preserved & frozen Bell loss, two injected
offsets, no post-validation reselection & preregistered IBM Kingston
end-to-end experiment; noninferior for both offsets\\
Decision procedure transports across tested stacks & same declared orbit and
role definitions; devicewise analysis & accept/reject/defer signature across
three IBM processors; numerical gaps are not pooled\\
Programmed $k=1,3$ signatures survive & frozen gate-level family and bases &
IQM two-block and IBM multi-device data\\
Strong/false/weak IBM roles separate & frozen roles and counts; retrospective
exact-binomial intervals & simultaneous 95\% reanalysis on Kingston and Fez\\
Native cubic pulse law & pulse-amplitude law measured & not tested\\
Universal saving & device-independent gap and cost & not claimed\\
\bottomrule
\end{tabular}
\end{table*}
The theorem and experiments have different roles. Algebra proposes a
quotient, statistics certifies or rejects it for one observed experiment, and
cost accounting decides whether to use it. The QPU data exercise acceptance,
rejection, deferral, replication, and shot-scaling signatures; they do not
replace the theorem assumptions or promote a gate-level construction to a
native pulse law.

\section*{Data and code availability}
The frozen manifests, count records, OpenQASM sources, analysis scripts, and
derived tables used for the reported results are organized in the project
reproduction packages, including the end-to-end archive identified above, and
are supplied as de-identified supplementary material.  No API credential,
account token, or provider-private metadata is included.  The exact
nonasymptotic certificate in \cref{app:proofs} and the prespecified bootstrap
analyses are supplied as distinct workflows.  If the supplementary archive is
deposited in a public repository, its persistent identifier will be reported
in the published data-availability statement.

\section*{Competing interests}
The authors declare no competing financial or non-financial interests.

\section*{Acknowledgments}
This work was initiated within the 2025 Texas State University HSMC program
and subsequently developed in the context of the Texas State University--SDT
memorandum of understanding.  The authors thank SDT for QuREKA access and
managed access to IQM Garnet.  Additional validation used IBM Quantum
services.  The theoretical claims and analysis remain the responsibility of
the authors.

\section*{Declaration of AI-assisted work}
OpenAI Codex was used in author-directed interactive sessions for analysis-code
drafting, algebraic and statistical consistency checks, organization of the
literature review, translation, and language editing.  It did not generate,
modify, or select QPU counts.  The authors checked the cited claims against the
linked publications, reran the disclosed analyses from frozen count files, and
reviewed every theorem, proof, numerical claim, and interpretation.  The
authors retain responsibility for the research questions, scientific
judgments, accuracy, originality, and final manuscript.  Session metadata and
the verification workflow are retained in the project record and can be
provided to an editor on request.

\begingroup
\small
\bibliographystyle{unsrturl}
\bibliography{references}
\endgroup
\end{document}